\documentclass[11pt]{article}\pdfoutput=1
\usepackage{cite}
\usepackage{amsmath,amsfonts,amssymb}
\usepackage[small,bf,hang]{caption}
\usepackage{slashed}
\usepackage{amsmath}
\usepackage{latexsym,epsfig}
\usepackage{arydshln}
 
\def\hybrid{
        \topmargin -20pt
        \oddsidemargin 0pt
        \headheight 0pt \headsep 0pt
        \textwidth 6.25in 
        \textheight 9.5in 
        \marginparwidth .875in
        \parskip 5pt plus 1pt \jot = 1.5ex}

\hybrid

 \csname
@addtoreset\endcsname{equation}{section}

\def\moth{\mathsurround=0pt}
\newdimen\zo \zo=0pt

\def\tick{\leaders\hrule height 0.5ex depth 0pt \hskip 0.5pt}
\def\upboxfill{$\moth \setbox\zo\hbox{\tick}%
  \hskip 3pt\hbox to 0pt{$\tick$\hss}\hrulefill \hbox to 7.5pt{$\tick$\hss}$}

\def\dtick{\leaders\hrule height .34pt depth 0.5ex \hskip 0.5pt}
\def\downboxfill{$\moth \setbox\zo\hbox{\dtick}%
  \hskip 2pt\hbox to 0pt{$\dtick$\hss}\hrulefill \hbox to 2pt{$\dtick$\hss}$}

\def\bec{\begin{center}}
\def\ec{\end{center}}

\def\be{\begin{equation}}
\def\ee{\end{equation}}
\def\bea{\begin{eqnarray}}
\def\eea{\end{eqnarray}}
\def\ba{\begin{array}}
\def\ea{\end{array}}

\def\ft#1#2{{\textstyle{{\scriptstyle #1}
\over {\scriptstyle #2}}}}

\begin{document}

\begin{titlepage}
\rightline{}
\rightline{June 2018}
\begin{center}
\vskip 2cm
{\Large \bf{The dual graviton in duality covariant theories}
}\\
\vskip 2.2cm

{\large\bf {Olaf Hohm${\,}^1$ and Henning Samtleben${\,}^2$}}
\vskip 1.6cm
{\it ${}^1$
Simons Center for Geometry and Physics, Stony Brook University,\\
Stony Brook, NY 11794-3636, USA}\\
ohohm@scgp.stonybrook.edu
\vskip .2cm

{\it ${}^2$ Univ Lyon, Ens de Lyon, Univ Claude Bernard, CNRS,\\
Laboratoire de Physique, F-69342 Lyon, France} \\
{henning.samtleben@ens-lyon.fr}

\end{center}

\bigskip\bigskip
\begin{center} 
\textbf{Abstract}

\end{center} 
\begin{quote}

We review (and elaborate on) the `dual graviton problem' in the context of duality covariant formulations 
of M-theory (exceptional field theories). These theories require fields that are interpreted as
components of the dual graviton in order to build complete multiplets under the exceptional 
groups E$_{d(d)}$, $d=2,\ldots, 9$. 
Circumventing no-go arguments, consistent theories for such fields have been constructed 
by incorporating additional covariant compensator fields. The latter 
fields are needed for gauge invariance but also for on-shell equivalence with $D=11$ and type IIB supergravity. 
Moreover, these fields play a non-trivial role in generalized Scherk-Schwarz 
compactifications, as for instance in consistent Kaluza-Klein truncation on  AdS$_4\times {\rm S}^7$.

\end{quote} 
\vfill
\setcounter{footnote}{0}
\end{titlepage}

\tableofcontents


\section{Introduction}

In string theory and supergravity it is often convenient or even necessary to pass from certain field variables 
to their `Poincar\'e duals'. For differential $p$-forms with suitable couplings this can be done straightforwardly by 
introducing a master action that treats the $(p+1)$-form field strength $F_{p+1}$ as an independent field 
whose Bianchi identity ${d}F_{p+1}=0$ is imposed by a Lagrange multiplier field, 
viewed as a differential 
form $A_{D-p-2}$. Upon integrating out $F_{p+1}$ one obtains an on-shell equivalent action 
for the dual $(D-p-2)$-form. Such duality transformations are instrumental, first, in order to 
describe the world-volume 
actions of certain branes and, second, 
to realize the U-duality symmetries of M-theory (and its low-energy actions) arising in toroidal compactifications  
in a manifestly local and covariant formulation.

The natural question arises whether theories of more complicated tensor fields have such 
`dual' formulations. Gravity linearized about flat space, i.e., the free massless spin-2 
theory for a symmetric second-rank tensor $h_{\mu\nu}$ on $D$-dimensional Minkowski space, permits a dual formulation in terms of a mixed Young-tableaux field 
$C_{\mu_1\ldots \mu_{D-3},\nu}$ --- the `dual graviton' \cite{Hull:2000zn,West:2001as}. 
This formulation can be obtained by 
essentially the same procedure as for $p$-forms, passing to a master action and then integrating out  
auxiliary fields. The existence of this dual formulation of linearized gravity has led to 
numerous speculations that the dual graviton (and more general mixed Young tableaux fields) 
may play an important role in the search for the elusive fundamental formulation of 
string/M-theory \cite{Hull:2000zn,West:2001as,Hull:2001iu,Nicolai:2005su}. 
Specifically, it has been suggested that, among other reasons, 
the dual graviton may be needed for a proper description of 
Kaluza-Klein monopole solutions of string theory \cite{Eyras:1998hn} 
and to realize enhanced U-duality-type symmetries. 
There are, however, strong no-go theorems excluding the existence of a manifestly covariant and local 
formulation of mixed Young tableaux fields beyond linear order \cite{Bekaert:2002uh,Bekaert:2004dz}.

In order to understand the significance of these no-go theorems, 
it is worthwhile to pause here for a moment and to reflect about what exactly the issue is. 
The issue is not to find a new \textit{physical} theory for a dual graviton field, 
because (non-linear) general relativity in physical or light-cone gauge is indistinguishable 
from a hypothetical theory of the dual graviton in light-cone gauge. 
Both would be formulated in terms of a symmetric tensor $\gamma_{ij}$ under 
the little group $SO(D-2)$ \cite{Goroff:1983hc} --- here one uses that the dual graviton 
$C_{i_1\ldots i_{D-3},j}$ can be replaced by $\gamma_{ij}$
by means of the $SO(D-2)$ epsilon symbol. Indeed, 
graviton and dual graviton are supposed to encode the same physical content.\footnote{
The name `dual graviton' is hence somewhat of a misnomer as it agrees with the 
`graviton' as usually defined: a massless spin-2 state whose interactions are governed 
by Einstein gravity (plus possible higher order corrections) 
no matter what field variables are used.} 
The real issue is rather one of a suitable \textit{formulation}, namely one that is non-linear, 
manifestly local and covariant, 
with a gauge symmetry so that \textit{i)} linearizing about flat space one recovers the free action 
of the dual graviton; and \textit{ii)} in light-cone gauge it is equivalent to general relativity. 
It is the existence of such a formulation that is excluded by the 
no-go theorems of \cite{Bekaert:2002uh,Bekaert:2004dz}.\footnote{More precisely, 
the no-go theorems of  \cite{Bekaert:2002uh,Bekaert:2004dz} 
exclude the existence of a non-linear theory that is manifestly local and covariant and  
satisfies  requirement 
\textit{i)}, irrespective of whether such a theory would be equivalent to general relativity and hence also satisfy 
requirement \textit{ii)}.}

More generally, as far as we can tell, 
there is no sharp \textit{physical} problem whose solution would require 
a theory of dual gravity as defined above. Nevertheless, it is reasonable to ask whether 
there are reformulations of (super-)gravity that feature a dual graviton-type field 
and that are useful for particular applications. It is indeed possible (in a surprisingly trivial fashion)  
to formulate general relativity so that it contains the dual graviton together with the 
usual graviton and a compensator gauge field \cite{West:2002jj,Boulanger:2008nd,Bergshoeff:2009zq}. 
This formulation is such that linearization about 
flat space yields, depending on a gauge choice, either standard linearized gravity or dual gravity, but 
this still begs the question what such a formulation is good for.  
There is one (reasonably 
sharp) \textit{mathematical} problem that has a bearing on the dual graviton issue, namely
the problem of finding a formulation of, say, $D=11$ supergravity 
that is \textit{duality covariant} under U-duality groups such as E$_{8(8)}$. 
In the following we outline how this problem arises and how it is resolved in exceptional field theory.

The E$_{8(8)}$ U-duality symmetry arises upon torus compactification of $D=11$ or 
type IIB supergravity to three dimensions. It is a non-linearly realized global symmetry, with the 
physical bosonic degrees of freedom being organized in a symmetric matrix ${\cal M}_{MN}$ parametrizing 
the coset space E$_{8(8)}/{\rm SO}(16)$.  Upon decomposing E$_{8(8)}$  w.r.t.~GL$(8)$ one may parametrize 
${\cal M}_{MN}$ in terms of (scalar) components, which include in particular fields  $\varphi_m$, $m=1,\ldots, 8$, 
that  are the on-shell duals of the Kaluza-Klein vector fields $A_{\mu}{}^{m}$ originating from the metric in $D=11$. 
As such, one may think of $\varphi_m$ as originating from the dual graviton, but as long as we are 
strictly in three dimensions there is no dual graviton problem. 
The problem arises if one attempts to formulate $D=11$ supergravity \textit{prior} to compactification 
in an E$_{8(8)}$ covariant way, as is done in exceptional field theory (ExFT). 
Here one decomposes all tensor fields and their indices as in Kaluza-Klein compactifications,
 but without truncating the 
coordinate dependence. The goal is then to reorganize the fields into duality covariant objects such as 
`scalars' ${\cal M}_{MN}$, vectors ${\cal A}_{\mu}{}^{M}$ and, more generally, higher tensors.  
The vector fields ${\cal A}_{\mu}{}^{M}$ generally contain the Kaluza-Klein vectors
$A_{\mu}{}^{m}$, and so for the E$_{8(8)}$ theory 
the question arises of how the eight components $\varphi_m$ of ${\cal M}_{MN}$ 
should be interpreted, in particular how such a theory should be matched with $D=11$ supergravity, 
which does not contain a dual graviton. (One could introduce a dual graviton, using the formulation of 
\cite{Boulanger:2008nd,Bergshoeff:2009zq}, but it should definitely be possible to match $D=11$ supergravity in the \textit{standard}
formulation.)

The resolution of the apparent conflict hinges on additional gauge fields and their associated gauge 
symmetries, which guarantee the correct counting of degrees of freedom and which are precisely as 
needed for the match with $D=11$ supergravity. In order to explain this we recall that in ExFT the fields depend on 
`external' coordinates $x^{\mu}$ and extended `internal' coordinates $Y^M$, subjects to `section constraints' 
for their dual derivatives of the form 
 \be\label{secconstrINTRO}
  \mathbb{P}_{MN}{}^{KL}\partial_K\otimes \partial_L \ = \ 0\;. 
 \ee 
Here, $\mathbb{P}$ is the projector onto suitable sub-representations in the tensor product 
of the fundamental representation (labelled by indices $M,N,\ldots$) with itself. This constraint should be 
interpreted in the sense that for any fields $A, B$ we set 
$\mathbb{P}_{MN}{}^{KL}\partial_K\partial_LA  =  0$ and 
$\mathbb{P}_{MN}{}^{KL}\partial_KA \,\partial_LB  = 0$. 
There are also generalized diffeomorphisms of the internal and external coordinates, which are consistent 
(obeying closure relations) thanks to the constraints (\ref{secconstrINTRO}). 
For the E$_{8(8)}$ theory their gauge parameters are $\Lambda^M, \Sigma_M$, but importantly 
the latter parameter needs to be \textit{covariantly constrained} in the sense that it satisfies 
constraints of the same type as the derivatives, 
  \be\label{secconstrINTROII}
  \mathbb{P}_{MN}{}^{KL}\partial_K\otimes \Sigma_L \ = \ 0\;, \quad {\rm etc}\,.
 \ee 
This constraint implies that the $\Sigma_M$ feature significantly fewer components 
than 248, with the precise non-vanishing field components depending on the `dual' choice of 
non-trivial coordinates among the $Y^M$. 
The vector fields are gauge fields for the internal (generalized) diffeomorphisms, 
and so we have a doubled set of gauge vectors ${\cal A}_{\mu}{}^{M}, {\cal B}_{\mu M}$, 
with the latter satisfying similar constraints as (\ref{secconstrINTROII}). 
Upon solving the section constraint, say as appropriate for $D=11$ supergravity, the $\Sigma_M$ gauge 
symmetries reduce to eight St\"uckelberg symmetries with parameters $\Sigma_m$, which are precisely 
sufficient in order to render the dual graviton components $\varphi_m$ pure gauge, thereby restoring the 
proper counting of degrees of freedom. (The same holds true for the type IIB solution of the constraint.)

In the remainder of this review we explain this resolution of the dual graviton problem in more technical 
detail for different ExFTs and how it is useful for applications such as using generalized Scherk-Schwarz compactifications for the consistency proofs of non-toroidal Kaluza-Klein truncations. 
Specifically, in sec.~2 we review the dualization of linearized gravity and explain the 
compensator formulation of \cite{Boulanger:2008nd,Bergshoeff:2009zq} that describes full, non-linear 
(super-)gravity. Moreover, we discuss the dimensional reduction of the dual graviton, 
which sets the stage for our subsequent discussion of ExFT, where certain dimensionally reduced 
components of the dual graviton are visible. In sec.~3 we review ExFT,  with a particular emphasis on 
how the components of the dual graviton enter the ExFT $p$-forms and induce additional compensating fields
with their associated gauge transformations. We illustrate this in detail for E$_{7(7)}$ and E$_{8(8)}$ ExFT. 
Finally, in sec.~4 we discuss the fate of the dual graviton in consistent Kaluza-Klein truncations
on non-trivial backgrounds.
Within ExFT these truncations are described as generalized Scherk-Schwarz reductions whose consistency 
requires the dual graviton and its compensating gauge field to be included. 
We close with a summary and outlook in sec.~5.

\section{Linearized dual gravity and its dimensional reduction } 

In this section we review the dualization of linearized gravity and discuss a reformulation of 
general relativity in terms of a dual graviton, the original graviton and a compensator gauge field. 
In the second subsection we briefly discuss the dimensional reduction of the dual graviton.

\subsection{Dual gravity in linearized and compensator form}

We begin with the frame-formulation of the Einstein-Hilbert action, in a form that is quadratic 
in first derivatives. It is written in terms of the coefficients of anholonomy 
 \bea\label{coeffanholo}
  \Omega_{ab}{}^{c} \ = \
  e_{a}{}^{\mu}e_{b}{}^{\nu}\left(\partial_{\mu}e_{\nu}{}^{c}-\partial_{\nu}e_{\mu}{}^{c}\right)\;, 
 \eea
where $e_{\mu}{}^{a}$ is the $D$-dimensional vielbein, as 
 \bea\label{EH}
  S_{\rm EH} \ = \ -\int d^D x\hspace{0.1em}e\left(\Omega^{abc}\Omega_{abc}+
  2\,\Omega^{abc}\Omega_{acb}-4\,\Omega_{ab}{}^b\Omega^{ac}{}_{c}\right)\;. 
 \eea
We next pass to a first-order formulation by
introducing an auxiliary field  $Y_{ab|c}$ that is antisymmetric in its first two indices 
but otherwise lives in a reducible representation of the Lorentz group \cite{West:2001as},
 \bea\label{first}
  S[Y,e] \ = \ -2\int d^Dx\hspace{0.1em}e
  \left(Y^{ab|c}\Omega_{abc}-\ft12Y_{ab|c}Y^{ac|b}+\tfrac{1}{2(D-2)}Y_{ab|}{}^{b}Y^{ac|}{}_{c}\right)\;.
 \eea
To show the equivalence to (\ref{EH}) we use that 
the field equation of $Y$ can be used to solve for $Y$ in terms of
$\Omega$,
 \bea\label{Ysol}
  Y_{ab|c} \ = \
  \Omega_{abc}-2\,\Omega_{c[ab]}+4\,\eta_{c[a}\Omega_{b]d}{}^{d}\;.
 \eea
Upon back-substitution into (\ref{first}) one recovers the Einstein-Hilbert action (\ref{EH}).

In order to obtain the dual formulation it is convenient to 
first rewrite
the action in terms of the Hodge dual of $Y^{ab|c}$,
 \bea\label{dualY}
  Y^{ab|c} \ \equiv \ \tfrac{1}{(D-2)!}\epsilon^{abc_1\cdots c_{D-2}}
  Y_{c_1\cdots c_{D-2}|}{}^{c}\;, 
 \eea
which yields 
\bea\label{firstdual}
 \begin{split}
  S \ = \ -\tfrac{2}{(D-2)!}\int
  d^Dx\hspace{0.1em}e\Big(&\epsilon^{abc_1\ldots
  c_{D-2}}Y_{c_1\ldots c_{D-2}|}{}^{c}\Omega_{abc}+\tfrac{D-3}{2(D-2)}
  Y^{c_1\ldots c_{D-2}|b}Y_{c_1\ldots
  c_{D-2}|b}\\
  &-\tfrac{D-2}{2}Y^{c_1\ldots c_{D-3}a|}{}_{a}Y_{c_1\ldots
  c_{D-3}b|}{}^{b}+\tfrac{1}{2}Y^{c_1\ldots c_{D-3}a|b}Y_{c_1\ldots
  c_{D-3}b|a}\Big)\;.
 \end{split}
\eea 
We then linearize this action about flat space, by writing for the frame field 
$e_{\mu}{}^{a}=\delta_{\mu}{}^{a}+\kappa\, h_{\mu}{}^{a}$,  where the
field $h_{\mu\nu}$ has no a priori symmetry.
By means of the background frame field given by the Kronecker delta,  
flat and curved indices can be identified. The coefficients of anholonomy (\ref{coeffanholo}) 
to first order in fluctuations can then be written as  
$\Omega_{\mu\nu\rho}=2\,\partial_{[\mu}h_{\nu]\rho}\,$. 
We next eliminate the graviton $h_{\mu\nu}$ in favor of a dual graviton, by noting that 
the field equation for
$h_{\mu\nu}$ is
 \bea\label{inte}
  \partial_{[\mu_1}Y_{\mu_2\ldots\mu_{D-1}]|\nu} \ = \ 0\;.
 \eea
The Poincar\'e lemma then implies that $Y$ is the curl of a
potential $C_{\mu_1\ldots\mu_{D-3}|\nu}$ (the `dual graviton') 
that is completely antisymmetric in its first $D-3$ indices,
  $Y_{\mu_1\ldots\mu_{D-2}|\nu} = 
  \partial_{[\mu_1}C_{\mu_2\ldots\mu_{D-2}]|\nu}$.
Inserting this back into (\ref{firstdual}) one obtains the 
action $S[C]$ for the dual graviton.

Let us now discuss the properties of the dual action in more detail.  
First, this action is of a general form discussed by Curtright \cite{Curtright:1980yk}. Defining the field strength 
 \bea\label{fieldstrength}
  F_{\mu_1\cdots\mu_{D-2}|\nu} \ \equiv \
  \partial_{[\mu_1}C_{\mu_2\cdots\mu_{D-2}]|\nu}\;, 
 \eea
the Curtright action is given by 
\begin{eqnarray}\label{Caction}
\begin{split}
  {\cal L}_{\rm C}(F) \ = & \ \frac{D-3}{2(D-2)} \,F^{\mu_1\cdots\mu_{D-2}|\nu}F_{\mu_1\cdots\mu_{D-2}|\nu}
  -\ft12(D-2)\, F^{\mu_1\cdots\mu_{D-3}\rho|}{}_{\rho}
  F_{\mu_1\cdots\mu_{D-3}\lambda|}{}^{\lambda} \\
   & \hspace*{.3cm} + \tfrac{1}{2}F^{\mu_1\cdots\mu_{D-3}\nu|\rho}
  F_{\mu_1\cdots\mu_{D-3}\rho|\nu}\;.
\end{split}
\end{eqnarray}
This action has the following gauge symmetries. First, 
the field strength (\ref{fieldstrength}) and hence the action are invariant under `dual diffeomorphisms' 
 \be\label{DUALDIff}
  \delta_{\Sigma} C_{\mu_1\ldots \mu_{D-3}|\nu} \ = \ \partial_{[\mu_1}\Sigma_{\mu_2\ldots  \mu_{D-3}]|\nu}\;. 
 \ee
Second, one may
check by an explicit computation that the action
$S[C]$ with Lagrangian (\ref{Caction})  
is invariant under the following St\"uckelberg symmetry~\cite{Boulanger:2003vs}
 \bea\label{stuckel}
  \delta_{\Lambda}C_{\mu_1\ldots\mu_{D-3}|\nu} \ = \
  -\Lambda_{\mu_1\ldots\mu_{D-3}\nu}\;,
 \eea
with completely antisymmetric shift parameter. 
From the point of view of the master action (\ref{firstdual}) (that is equivalent to the Einstein-Hilbert action) 
this is simply a consequence of the local Lorentz symmetry that to first order acts on the fluctuation as 
$\delta_{\Lambda} h_{\mu\nu}\propto \Lambda_{\mu\nu}$. 
Consequently, the totally antisymmetric part of $C_{\mu_1\ldots\mu_{D-3}|\nu}$ can be
gauge-fixed to zero inside $S[C]\,$, in the same way that the antisymmetric part of $h_{\mu\nu}$ can 
be gauged away in standard linearized gravity. (It should be emphasized, however, that in the master action (\ref{firstdual}) 
we cannot gauge away 
the antisymmetric parts of $h$~\cite{Boulanger:2003vs}.) 

The field $D_{\mu_1\ldots\mu_{D-3}|\nu}$ obtained by gauging away the totally antisymmetric part of $C$ 
carries a specific Young-diagram symmetry. 
The characteristics of such mixed
Young tableaux fields have been studied  in
\cite{Curtright:1980yk,Aulakh:1986cb}, where it has been shown that they transform
under two types of gauge transformations as follows 
 \bea\label{dualgauge}
  \delta
  D_{\mu_1\cdots\mu_{D-3}|\nu}=
   \partial_{[\mu_1}\alpha_{\mu_2\cdots\mu_{D-3}]|\nu}
  +\partial_{[\mu_1}\beta_{\mu_2\cdots\mu_{D-3}]\nu}
   -(-1)^{D-3}\partial_{\nu}\beta_{\mu_1\cdots\mu_{D-3}}\;, 
 \eea
where $\alpha$ lives in the $(D-4,1)$ Young tableau, and
$\beta$ is completely antisymmetric. 
These transformations originate after gauge fixing 
from (\ref{DUALDIff}) and compensating local Lorentz transformations (\ref{stuckel}).  
The field strength (\ref{fieldstrength}), with the $C$ field replaced by the $D$ field, is invariant under 
$\alpha$-transformations, which therefore are a manifest invariance of the Curtright action. 
In contrast, the $\beta$ transformations are a non-manifest invariance. 
Indeed, the relative coefficients can be fixed by requiring gauge invariance under
$\beta$-transformations.

\medskip

After having discussed the dualization of linearized gravity, let us now return to full non-linear 
Einstein gravity. We first note that the above dualization procedure cannot be applied 
to the non-linear theory, because the field equations for $h_{\mu\nu}$ no longer imply that the curl of $Y$ vanishes, c.f.~(\ref{inte}). Rather, one obtains an equation of the schematic form $\partial Y\sim Y^2$, which does
not imply that $Y$ can be written as the curl of a dual graviton field. 
This obstacle for extending the dualization to the non-linear level is in perfect agreement with the no-go theorems 
of \cite{Bekaert:2002uh,Bekaert:2004dz}  that prohibit the existence of a manifestly covariant and local 
dual formulation. Instead, we will discuss now a non-linear formulation that features both the graviton and dual graviton together with a compensating gauge field  \cite{West:2002jj,Boulanger:2008nd}. This theory is a straightforward re-interpretation of 
the master action (\ref{firstdual}) that, however, turns out to be quite prescient for the exceptional 
field theory formulations to be discussed below. 

This theory is obtained by starting from the quadratic Curtright action (\ref{Caction}) and coupling it to 
dynamical gravity described by a vielbein field $e_{\mu}{}^{a}$. This yields the 
`covariantized' Curtright Lagrangian
 \begin{eqnarray}\label{covCurt}
   {\cal L}_C(e,{F}) &=& \tfrac{D-3}{2(D-2)}\,e\,
   {F}^{\mu_1\ldots\mu_{D-2}|a}{F}_{\mu_1\ldots\mu_{D-2}|a}
   -\tfrac{D-2}{2}\;e\, e_{\nu}{}^{a}\,e_{b}{}^{\rho}\,
   {F}^{\mu_1\ldots\mu_{D-3}\nu|}{}_{a}\,
   {F}_{\mu_1\ldots\mu_{D-3}\rho|}{}^{b}
    \nonumber \\
   && +\tfrac{1}{2}\,e\,e_{\nu}{}^{b}\,e_{a}{}^{\rho}\,
   {F}^{\mu_1\ldots\mu_{D-3}\nu|a}\,{F}_{\mu_1\ldots\mu_{D-3}\rho|b}
   \;, 
 \end{eqnarray}
which is now fully diffeomorphism invariant. Here we interpret  the dual graviton 
$C_{\mu_1\ldots\mu_{D-3}\,a}$ as a $(D-3)$-form in the vector representation of the Lorentz group,  
whose field strength $F_{[D-2]a}$ is defined as in (\ref{fieldstrength}). The action then admits 
a (still abelian) dual diffeomorphism symmetry that acts on $C_{\mu_1\ldots\mu_{D-3}\,a}$ as an 
ordinary $p$-form gauge symmetry (with a $(D-4)$-form gauge parameter in the vector representation 
of the Lorentz group).   
However, the (dual) local Lorentz transformations (\ref{stuckel}) are no longer an invariance of the action, 
and hence (\ref{covCurt}) is not fully consistent. 
In order to repair this, we have to modify the field strength by adding a compensating gauge field in the form of 
a St\"uckelberg coupling,
 \bea\label{shiftfield}
  \hat{F}^{\qquad\quad ~a}_{\mu_1\cdots\mu_{D-2}} \ \equiv \
  \partial_{[\mu_1}C_{\mu_2\cdots\mu_{D-2}]}{}^a \ + \ 
  Y^{\qquad\quad ~a}_{\mu_1\cdots\mu_{D-2}}\;. 
 \eea
The field $Y$ is now interpreted as a compensating gauge field for  St\"uckelberg gauge symmetries 
that act as 
 \bea\label{shiftsym}
  \delta Y^{\qquad\quad ~a}_{\mu_1\cdots\mu_{D-2}} \ = \
  \partial^{}_{[\mu_1}\Sigma_{\mu_2\cdots\mu_{D-2}]}{}^{a}\;,
  \qquad
  \delta C_{\mu_1\cdots\mu_{D-3}}{}^{a} \ = \
  -\Sigma_{\mu_1\cdots\mu_{D-3}}{}^{a}\;, 
 \eea
and that hence leave the field strength $\hat F$ invariant.

Since we coupled the free Curtright action to the dynamical gravity field $e_{\mu}{}^{a}$ it would be natural 
to add a kinetic Einstein-Hilbert term to (\ref{covCurt}). However, this would lead to a doubling of the gravity 
degrees of freedom, instead of a reformulation of Einstein gravity, and it would also not restore the local 
Lorentz invariance. The correct procedure is instead to add a topological term that couples $e_{\mu}{}^{a}$
to the compensating shift gauge field $Y$. We thus consider the total action \cite{Boulanger:2008nd} 
\bea\label{nonlin}
  S[e,C,Y] &=& \int d^Dx \Big[ {\cal L}_C(e,\hat{F})
   \ + \ 2\kappa^{-1}\,\varepsilon^{\mu_1\ldots\mu_{D-2}\nu\rho}\,
  Y_{\mu_1\ldots\mu_{D-2}\,a}\;\partial_{\nu}e_{\rho}{}^{a} \Big]\;,
 \eea
where we restored Newton's constant $\kappa$. 
Let us verify that the theory defined by this action is equivalent to Einstein gravity. 
To this end we note that the St\"uckelberg shift symmetry (\ref{shiftsym}) can be gauge fixed by setting 
$C=0$ so that $\hat{F}=Y$, in which case (\ref{nonlin}) reduces to the original master action (\ref{firstdual}). 
As the latter leads to the Einstein-Hilbert action upon integrating out $Y$, we have shown that (\ref{nonlin})
is equivalent to Einstein's general relativity. 
On the other hand, we may linearize about flat space before gauge fixing and/or integrating out fields. 
In this case, the field equation of $e_{\mu}{}^{a}$ reduces to $dY=0$, which in turn implies that the
St\"uckelberg symmetry (\ref{shiftsym}) can be gauge fixed by setting $Y=0$. The action (\ref{nonlin})
then reduces to the Curtright action (\ref{Caction}). The action (\ref{nonlin}) thus provides a universal formulation 
of gravity that 
features both the graviton and dual graviton, together with a compensating gauge field. 
Although this formulation is a minor extension of the original master action (\ref{firstdual}), it turns out that 
its basic mechanism of compensating gauge fields is realized, in a more subtle and duality covariant version, 
in exceptional field theory.  

It is instructive to investigate the field equations following from the above action, 
which take the form of first-order duality relations. 
Varying with respect to the gauge field $Y$ one finds
 \begin{eqnarray}\nonumber
  e^{-1}\varepsilon^{\mu_1\ldots\mu_{D-2}\nu\rho}\Omega_{\nu\rho}{}^{a}
  &=& -\frac{D-3}{D-2}\,\hat{F}^{\mu_1\ldots\mu_{D-2}|a} +(-1)^{D-3}
  (D-2)\,e_{\rho b}e^{a[\mu_1}\hat{F}^{\mu_2\ldots\mu_{D-2}]\rho|b} \\
  &&-(-1)^{D-3}e_{\rho}{}^{a}e_{b}{}^{[\mu_1}\hat{F}^{\mu_2\ldots\mu_{D-3}]\rho|b}\;, 
 \label{dualitygrav}
 \end{eqnarray}
while the field equation for $e_{\mu}{}^{a}$ reads 
 \bea\label{2ndduality}
  e^{-1}\varepsilon^{\mu\mu_1\ldots\mu_{D-1}}\partial_{\mu_1}
  Y_{\mu_2\ldots\mu_{D-1}|a}
  \ = \ \ft12 \,e^{-1}\frac{\delta{\cal L}_{\rm C}(e,\hat{F})}
  {\delta e_{\mu}{}^{a}}\;.
 \eea
These combined first-order equations imply the full non-linear Einstein equations, 
which is of course guaranteed from the definition of the master action. 
To this end one has to take suitable derivatives of (\ref{dualitygrav}) and use on the right-hand side 
the Bianchi identity of $\hat{F}$, which reads schematically $d\hat{F}^a=dY^a$. 
One can then use the second duality relation (\ref{2ndduality}) in order to eliminate $dY^a$. 
Alternatively, one may solve (\ref{dualitygrav}) for $Y$ in terms of $e$ and $C$ and 
then insert into (\ref{2ndduality}) upon which $C$ drops out and the Einstein equations are 
obtained. In particular, the first equation (\ref{dualitygrav}) by itself has no physical content 
(in contrast to the linear duality relation without compensating gauge field) in that it can be 
viewed as a mere definition of $Y$, but the point is that $Y$ in turn satisfies an equation, eq.~(\ref{2ndduality}), 
so that the combined system implies the dynamical second order Einstein equations. 
This mechanism of `hierarchical' duality relations is very natural for 
the tensor hierarchy structure in gauged supergravity \cite{Bergshoeff:2009ph} 
and ExFT and will recur 
in several places below. Finally, we note that here arbitrary matter couplings could be introduced 
by adding the matter action to (\ref{nonlin}), without modifying ${\cal L}_C$. 
This leaves
the first duality relation unchanged, but adds to the second duality
relation (\ref{2ndduality}) the standard energy-momentum tensor
$T^{\mu}{}_{a}\sim\delta{\cal L}_{\rm M}/\delta e_{\mu}{}^{a}$,
which in turn re-appears in the Einstein equation in the usual way.

\subsection{Dimensional reduction of dual gravity} 

We will now discuss some aspects of the dimensional or Kaluza-Klein reduction of theories involving the 
dual graviton. In principle, we could work out the reduction of the full non-linear master action (\ref{nonlin}) 
including all fields, but here we content ourselves with a more schematic discussion of the type of 
fields appearing in lower dimensions. This sets the stage for our subsequent discussion of related 
fields in exceptional field theory. It is then sufficient to inspect the linearized theory, and 
here it is convenient to work with the master action in the form (\ref{first}).
In order to distinguish between world indices in different dimensions we will temporarily change 
notation and denote full $D$-dimensional spacetime indices by $\hat{\mu}, \hat{\nu},\ldots=0,\ldots, D-1$, 
so that the (linearized) master action reads 
 \bea\label{firstsecond}
  S[Y,e] \ = \ -2\int d^Dx\hspace{0.1em}e
  \left(Y^{\hat{\mu}\hat{\nu}|\hat{\rho}}\,\Omega_{\hat{\mu}\hat{\nu}\hat{\rho}}
  -\ft12Y_{\hat{\mu}\hat{\nu}|\hat{\rho}}Y^{\hat{\mu}\hat{\rho}|\hat{\nu}}
  +\tfrac{1}{2(D-2)}Y_{\hat{\mu}\hat{\nu}|}{}^{\hat{\nu}}Y^{\hat{\mu}\hat{\rho}|}{}_{\hat{\rho}}\right)\;, 
 \eea
where 
 \be\label{linCOeff}
  \Omega_{\hat{\mu}\hat{\nu}\hat{\rho}} \ = \ \partial_{\hat{\mu}}h_{\hat{\nu}\hat{\rho}}
  -\partial_{\hat{\nu}}h_{\hat{\mu}\hat{\rho}}
 \ee 
are the linearized coefficients of anholonomy. This action is invariant under (linearized) local Lorentz transformations, 
with $Y$ transforming as 
 \be\label{LinLor}
  \delta_{\Lambda}Y_{\hat{\mu}\hat{\nu}|\hat{\rho}} \ = \ -2\,\partial_{\hat{\rho}}\Lambda_{\hat{\mu}\hat{\nu}}
  -4\,\eta_{\hat{\rho}[\hat{\mu}}\,\partial^{\hat{\sigma}}\Lambda_{\hat{\nu}]\hat{\sigma}}\;. 
 \ee 

We now perform the dimensional reduction of (\ref{firstsecond}) by decomposing the spacetime indices 
as $\hat{\mu}=(\mu, m)$, where $\mu=0,\ldots, n-1$ are the external indices and $m=1,\ldots, d=D-n$ 
are the internal indices, and assuming that all fields are independent of the internal coordinates, thus setting 
$\partial_m=0$. For the (linearized) frame field we then write 
\be\label{KKsplit}
 h_{\hat{\mu}\hat{\nu}} \ = \ \begin{pmatrix}   h_{\mu\nu} & A_{\mu m}\\
  h_{m\mu} & \phi_{m,n} \end{pmatrix}\;. 
\ee 
Note that usually one fixes the Lorentz gauge (partially) by setting $h_{m\mu}=0$, but in the present 
context it is important to keep all fields so that we do not lose field equations. 
We will confirm momentarily, however, that  $h_{m\mu}$ is non-propagating. 
Moreover, we recall that $h_{{\mu}{\nu}}$ and $\phi_{m,n}$
carry symmetric and antisymmetric parts. The surviving gauge symmetries are given in terms 
of these components by 
 \be
  \begin{split}
   \delta h_{\mu\nu} \ &= \ \partial_{\mu}\xi_{\nu} - \Lambda_{\mu\nu}\;, \\
   \delta A_{\mu m} \ &= \ \partial_{\mu} \xi_{m}-\Lambda_{\mu m}\;, \\
   \delta \phi_{m,n} \ &= \ -\Lambda_{mn}\;, \\
   \delta h_{m\mu} \ &= \ \Lambda_{\mu m}\;. 
  \end{split}
 \ee
The non-vanishing components of the coefficients of anholonomy (\ref{linCOeff}) are given by 
 \be
  \begin{split}
   \Omega_{\mu\nu\rho} \ &= \ 2\, \partial_{[\mu} h_{\nu]\rho}\;, \\
   \Omega_{\mu\nu m} \ &= \ F_{\mu\nu m} \ \equiv \ \partial_{\mu}A_{\nu m} - \partial_{\nu}A_{\mu m}\;, \\
   \Omega_{\mu mn} \ &= \ \partial_{\mu}\phi_{m,n}\;, \\
   \Omega_{\mu m\nu} \ &= \ \partial_{\mu} h_{m\nu}\;. 
  \end{split}
 \ee 
Integrating out $Y$ from (\ref{firstsecond}) naturally yields the free kinetic terms for 
the graviton $h_{\mu\nu}$, the Kaluza-Klein vector $A_{\mu m}$ and the Kaluza-Klein scalars $\phi_{mn}$, 
while the unphysical $h_{m\mu}$ drops out. (More precisely, for the vectors 
only the shift-invariant combination $A_{\mu m}+ h_{m\mu}$ enters the action, which can therefore be identified 
with the Kaluza-Klein vectors.)

In order to obtain the dual theory, we vary w.r.t.~the original Kaluza-Klein fields in (\ref{KKsplit}). 
To this end, we need the dimensional reduction of the $Y\Omega$ term in  (\ref{firstsecond}): 
 \be\label{YOmred}
  Y^{\hat{\mu}\hat{\nu}|\hat{\rho}}\,\Omega_{\hat{\mu}\hat{\nu}\hat{\rho}}
  \ = \ 2\, Y^{\mu\nu|\rho}\partial_{\mu} h_{\nu\rho} + Y^{\mu\nu|m} F_{\mu\nu m} 
  + 2\,Y^{\mu m|n}\partial_{\mu}\phi_{m,n} +2\, Y^{\mu m|\nu}\partial_{\mu}h_{m\nu}\;,  
 \ee
which contain the only couplings to the original Kaluza-Klein fields (\ref{KKsplit}). 
Thus, varying the action w.r.t.~the physical fields $h$, $A$ and $\phi$, respectively, yields 
 \be\label{detY}
  \begin{split}
   \partial_{\mu}Y^{\mu\nu|\rho} \ &= \ 0\qquad \Rightarrow \qquad Y^{\mu\nu|\rho} \ = \ 
   \partial_{\sigma}C^{\sigma\mu\nu|\rho}\;, \\
   \partial_{\mu}Y^{\mu\nu|m} \ &= \ 0\qquad \Rightarrow \qquad 
   Y^{\mu\nu|m} \ = \ \partial_{\rho}B^{\rho\mu\nu|m}\;, \\
   \partial_{\mu}Y^{\mu m|n} \ &= \ 0 \qquad \Rightarrow \qquad 
   Y^{\mu m|n} \ = \ \partial_{\nu}E^{\nu\mu m|n}\;, 
  \end{split}
 \ee  
 where we used the Poincar\'e lemma. (Here we use the convention that like-wise indices 
 that are not separated by a bar are assumed to be totally antisymmetric.) 
 Varying w.r.t.~the unphysical field $h_{m\mu}$ yields 
  \be\label{Kfields}
   \partial_{\mu}Y^{\mu m|\nu} \ = \ 0\qquad \Rightarrow \qquad
   Y^{\mu m|\nu} \ = \ \partial_{\rho}K^{\rho\mu m|\nu}\;. 
  \ee
 Upon reinserting into the action, all terms coming from the expansion of $Y\Omega$ then 
 reduce to total derivatives, while the $Y^2$ terms yield the proper kinetic terms for the 
 dual (generally propagating) fields $C$, $B$ and $E$.  
  
 The fields thus obtained can be written a little more suggestively 
 as 
  \be\label{dualGRAVcomppp}
   \begin{split}
    C_{[n-3,1]} \,&: \qquad \; C_{\mu_1\ldots \mu_{n-3}|\nu} \ \propto \ 
    \epsilon_{\mu_1\ldots \mu_{n-3} \rho\sigma\lambda} \,C^{\rho\sigma\lambda}{}_{|\nu}\;, \\
    B_{[n-3]m}\,&: \qquad B_{\mu_1\ldots \mu_{n-3}|m} \ \propto \  
    \epsilon_{\mu_1\ldots \mu_{n-3}\nu\rho\sigma} \,B^{\nu\rho\sigma}{}_{|m}\;,  \\
    E_{[n-2]m,n} \,&: \quad \;E_{\mu_1\ldots \mu_{n-2}|m,n} \ \propto \  
     \epsilon_{\mu_1\ldots \mu_{n-2}\nu\rho} \,E^{\nu\rho}{}_{|m,n}\;,
   \end{split}
  \ee  
 and are thus naturally identified with components of the dimensionally reduced dual graviton $C_{[D-3,1]}$. 
 Eliminating the $Y$ fields determined by (\ref{detY}) inside (\ref{firstsecond}) then yields the second-order 
 actions for the dual fields that one would also obtain by dimensionally reducing the Curtright action directly. 
Note that the $(n-3)$-forms $B_m$ and the $(n-2)$-forms $E_{m,n}$ are the standard duals of vectors 
and scalars, respectively. 
More precisely, a suitable combination of the $(n-3)$-forms determined in the second line of (\ref{detY}) 
and of the $(n-3)$-form sub-representations contained in the $K$ fields in (\ref{Kfields}) play the role 
of the duals to the vectors. The remaining sub-representations of the $K$-field are either non-propagating 
or pure gauge (noting that with (\ref{LinLor}) the local Lorentz transformations imply 
$\delta K_{\mu\nu\,m|\rho}=-4\,\eta_{\rho[\mu}\Lambda_{\nu]m}$, so that the trace part is pure gauge).  
Finally, the components of the $Y$ fields that have not yet been determined, which are 
  \be
   Y_{\mu m|n}\;, \quad Y_{mn|\mu}\;, \quad Y_{mn|k}\;, 
  \ee 
 then enter the action purely quadratically and hence are non-propagating and can be eliminated 
algebraically. Relatedly, we note that the naive decomposition of the dual graviton $C_{[D-3,1]}$ under the 
Kaluza-Klein split yields more component fields than are contained in (\ref{dualGRAVcomppp}), 
but it follows from the above analysis that these fields are non-propagating and can thus directly 
be eliminated from the action. 
 
Let us summarize the dual graviton components relevant in each dimension. 
In the exceptional field theory formulations to be discussed in the next section, the external graviton degrees 
of freedom will always be described conventionally.  
In the above formulation this corresponds to integrating out 
the external components $Y_{\mu\nu|\rho}$, which leads to the standard Einstein-Hilbert action. 
The dual graviton components needed in each dimension are those dual to the Kaluza-Klein vectors
and hence given for $D=3,4,5$ by 
 \be
  \begin{split}
   D \ = \ 3\,:\qquad B_m \qquad &\rightarrow \qquad 
   D_{\mu}B_m \ = \ \partial_{\mu}B_m  \ + \ \cdots  \ + \ \tilde{Y}_{\mu|m}\;, \\
   D \ = \ 4\,: \qquad B_{\mu m} \qquad &\rightarrow \qquad 
   F_{\mu\nu \,m} \ = \ 2\,\partial_{[\mu} B_{\nu]m} \ + \ \cdots \  + \ \tilde{Y}_{\mu\nu|m}\;, \\
   D \ = \ 5\,: \qquad B_{\mu\nu m} \qquad &\rightarrow \qquad  
   H_{\mu\nu\rho \,m }  \ = \ 3\,\partial_{[\mu} B_{\nu\rho]m}\ + \ \cdots \  + \ \tilde{Y}_{\mu\nu\rho|m}
   \;. 
  \end{split} 
  \label{DBF}
 \ee
Here we also indicated the compensating gauge fields entering through St\"uckelberg-type couplings 
in the compensator formulation that exists at the full non-linear level. These fields correspond, 
in dimensions $D=3,4,5$, to extra vectors, two-forms and three-forms, respectively, and are a re-interpretation 
(and dualization) of the $Y$ fields. 

In addition, also the dual graviton components dual to the Kaluza-Klein scalars will typically be visible below, 
which in dimensions $D=3,4,5$ are given by $E_{\mu \,m,n}$, $E_{\mu\nu \, m,n}$ and $E_{\mu\nu\rho \, m,n}$, 
respectively. Let us finally note that there is an intriguing interplay between these dual graviton components 
and those originating from the Kaluza-Klein vectors, which is already visible at linearized level 
provided one keeps the full coordinate dependence as in exceptional field theory. 
To illustrate this, let us return to the master action (\ref{firstsecond}) and again perform the Kaluza-Klein 
split (\ref{KKsplit}), but now without truncating the coordinate dependence. This leads to further 
terms in (\ref{YOmred}) involving the internal derivative $\partial_m$ and hence modifies the 
field equations of the Kaluza-Klein fields accordingly. In particular, 
the field equations for the Kaluza-Klein vectors are now solved by 
 \be 
   Y^{\mu\nu|m} \ = \ \partial_{\rho}B^{\rho\mu\nu|m} + \partial_nE^{\mu\nu\,n|m}\;. 
 \ee
(This equation can of course be obtained directly by solving the field equation for the full 
$h_{\hat{\mu}\hat{\nu}}$ as 
$Y^{\hat{\mu}\hat{\nu}|\hat{\rho}}=\partial_{\hat{\sigma}}C^{\hat{\sigma}\hat{\mu}\hat{\nu}|\hat{\rho}}$ and reading off this component.) Dualizing now as in (\ref{dualGRAVcomppp}), say in four external dimensions, this naturally leads to the field strength 
 \be
  F_{\mu\nu m}(B) \ = \ 2\,\partial_{[\mu}B_{\nu] m} \ + \ \cdots \ + \ \partial_n E_{\mu\nu\,m}{}^{n}  
  \  + \ \tilde{Y}_{\mu\nu|m}\;, 
  \label{Fm4}
 \ee
where we also included the compensating two-form, while the ellipsis again denotes terms that would arise in 
the full non-linear theory.\footnote{Upon redefining $E_{\mu\nu\,m}{}^{n}  \rightarrow E_{\mu\nu\,m}{}^{n}  +
\alpha\, E_{\mu\nu\,k}{}^{k}\,\delta_{m}{}^{n}$ we could change the structure of this term to resemble more closely 
some of the formulas below.} The terms displayed in this field strength follow naturally 
by Kaluza-Klein decomposing (\ref{shiftfield}) (where one has to recall that, as in (\ref{detY}), 
redefinitions with the external and internal epsilon symbols are needed), while the non-linear terms 
not displayed are generated by Kaluza-Klein redefinitions, c.f.~eqs.~(4.30) in \cite{Hohm:2013vpa}. 
The above field strength is invariant under gauge transformations with one-form 
parameters $\Sigma_{\mu\, m}{}^{n}$:
 \be
  \delta B_{\mu\,m} \ = \ -\partial_n\Sigma_{\mu \,m}{}^{n}\;, \qquad 
  \delta E_{\mu\nu\,m}{}^{n}\ = \ 2\,\partial_{[\mu}\Sigma_{\nu]m}{}^{n}\;,
 \ee
which is a direct consequence of the dual diffeomorphism symmetry, c.f.~(\ref{DUALDIff}).
Below we will repeatedly encounter this general structure of `hierarchical' gauge symmetries 
and their invariant field strengths.

\section{The dual graviton in exceptional field theory}

In this section, we review the structure of $p$-forms in exceptional field theories
and discuss explicitly the structure and couplings of $(8-d)$- and $(9-d)$-forms.
Among the latter feature the covariantly constrained compensator fields that are
required for the construction of covariant field strengths and closure of the gauge algebra.
Upon solving the section constraints and recovering the 11-dimensional field equations,
these forms carry components from the 11-dimensional dual graviton and compensator field,
respectively.
We first discuss the generic structure and then illustrate the results for E$_{7(7)}$ and
E$_{8(8)}$ ExFT.

\subsection{Exceptional field theory and the tensor hierarchy}

Exceptional field theory (ExFT) is a framework that embeds all  $D=11$ and $D=10$ supergravities 
in a way manifestly covariant under the E$_{d(d)}$ group that becomes a global symmetry 
after dimensional reduction~\cite{Hohm:2013pua,Hohm:2013vpa,Hohm:2013uia}.
More precisely, in ExFT fields fall into representations of E$_{d(d)}$ and coordinates 
split into $(11-d)$ external coordinates $\{x^\mu\}$ and internal coordinates $\{y^m\}$ of which the latter 
are embedded into the fundamental representation of E$_{d(d)}$ with the coordinate dependence restricted by 
the section constraints
\bea
Y^{MK}{}_{NL}\,\partial_M\otimes \partial_K &=& 0
\;.
\label{section}
\eea
Here, $\partial_M$ define derivatives w.r.t.\ coordinates $\{Y^M\}$ transforming in the fundamental
representation of E$_{d(d)}$, and $Y^{MK}{}_{NL}$ is a constant E$_{d(d)}$-invariant tensor, see~\cite{Berman:2012vc}.
Any solution to (\ref{section}) breaks E$_{d(d)}$ by restricting the coordinate dependence of all fields 
to $d$ or $(d-1)$ coordinates, whereupon the ExFT field equations reproduce the $D=11$ and IIB
field equations, respectively.

The ExFT field equations are manifestly invariant under generalized diffeomorphisms, acting as
\bea
{\cal L}^{[\lambda]}_\Lambda \, V^M &=& 
\Lambda^K \partial_K V^M + \kappa\, \mathbb{P}^M{}_N{}^K{}_L\,\partial_K\Lambda^L\,V^N
+\lambda\,\partial_N \Lambda^N V^M
\;,
\label{gen_diff}
\eea
on a vector field $V^M$ of internal weight $\lambda$. Here, $\kappa$ is a constant 
(fixed by closure of the algebra), and $\mathbb{P}^M{}_N{}^K{}_L$ denotes the 
projector onto the adjoint representation of E$_{d(d)}$,
explicitly given by
\bea
\kappa\,\mathbb{P}^M{}_N{}^K{}_L&=& \kappa\,(t_\alpha)_N{}^M\,(t^\alpha)_L{}^K
~=~
Y^{MK}{}_{LN}-\delta^M{}_L\,\delta^K{}_N-\lambda_d\,\delta^M{}_N\,\delta^K{}_L
\;,
\label{PY}
\eea
in terms of the E$_{d(d)}$ generators $(t_\alpha)_N{}^M$,
and related to the tensor $Y^{MK}{}_{NL}$
defining the section constraint (\ref{section}), with $\lambda_d\equiv\frac1{9-d}$\,.
We refer to the three contributions to (\ref{gen_diff}) as the transport term, the rotation term,
and the weight term, respectively.

In ExFT, the generalized diffeomorphisms (\ref{gen_diff}) are a local symmetry
w.r.t.\ the external coordinates, implemented by a gauge connection ${\cal A}_\mu{}^M$ 
and standard covariant external derivatives
\bea
D_\mu &=& \partial_\mu - {\cal L}_{{\cal A}_\mu}
\;.
\label{covder}
\eea
The remaining ExFT fields organize into E$_{d(d)}$ representations that are scalars and $p$-forms
w.r.t.\ the external coordinates. Accordingly the latter come with gauge transformations
with gauge parameters of rank $(p-1)$, defining a non-abelian tensor hierarchy on top
of the generalized diffeomorphisms (\ref{gen_diff}).
In particular, the components $B_m$ and $E_{m,n}$ (\ref{dualGRAVcomppp}) from the higher-dimensional 
dual graviton sit within $(8-d)$-forms ${\cal B}_M$ and $(9-d)$-forms ${\cal C}_\alpha$, in the fundamental and the adjoint 
representation of E$_{d(d)}$, respectively.

As a characteristic feature of non-abelian tensor hierarchies \cite{deWit:2008ta}, the covariant field strength
associated to a $p$-form gauge potential ${\cal C}^{[p]}$ is
of the schematic form
\bea
{\cal F}^{[p+1]} &=&
D {\cal C}^{[p]} + \dots + {\cal D} {\cal C}^{[p+1]}
\;.
\label{Fnonab}
\eea
Here, $D$ is the covariant external derivative (\ref{covder}), the ellipsis represents possible 
Chern-Simons-like contributions polynomial in lower-rank $p$-forms, and ${\cal D}$ is a differential
operator in the internal derivatives acting on a $(p+1)$-form gauge potential ${\cal C}^{[p+1]}$.
In standard non-abelian field theories, ${\cal D}$ is typically given by an algebraic operator and
describes the St\"uckelberg-type coupling of a higher-rank gauge potential.

Accordingly, the $p$-form potentials are subject to gauge transformations of the type
\bea
\delta_\Lambda {\cal C}^{[p]} &=& D\Lambda^{[p-1]} + \dots - {\cal D} \Lambda^{[p]}
\;,
\label{deltaC}
\eea
in order to leave the field strength (\ref{Fnonab}) invariant. Consistency (in particular closure
of the full gauge algebra) requires that the
internal derivative operator ${\cal D}$ 
defines a tensor under
generalized diffeomorphisms (in analogy to the exterior derivative $D$ in the external sector). 
As observed in~\cite{Cederwall:2013naa}, this holds true for $p\le 7-d$ but fails at $p=8-d$, where
the components of the dual graviton first enter the ExFT fields.
Let us make this explicit. The $(8-d)$-forms ${\cal B}_M$ in ExFT transform in the (dual) fundamental
representation of E$_{d(d)}$, and the gauge transformations (\ref{deltaC}) take the explicit form
\bea
\tilde\delta_\Lambda {\cal B}_M &=& D\Lambda_M + \dots - (t^\alpha)_M{}^N\,\partial_N \Lambda_\alpha
\;,
\label{deltaBexp}
\eea
with the $(8-d)$-form gauge parameter $\Lambda_\alpha$ in the adjoint representation of E$_{d(d)}$,
and the operator ${\cal D}$ explicitly realized in terms of the E$_{d(d)}$ generators $(t^\alpha)_M{}^N$.
The notation $\tilde\delta$ indicates that (\ref{deltaBexp}) is not the final answer.
Using (\ref{gen_diff}) and (\ref{PY}) it is  a straightforward exercise that given a tensor $\Lambda_\alpha$
transforming as
\bea\label{deltaLambda}
\delta_\Lambda \Lambda_\alpha &=& 
{\cal L}_\Lambda^{(\lambda)} \Lambda_\alpha ~\equiv~
\Lambda^K \partial_K \Lambda_\alpha + \kappa\, 
f_{\beta\alpha}{}^\gamma\,(t^\beta)_L{}^K \,\partial_K\Lambda^L\,\Lambda_\gamma
+\lambda\,\partial_N \Lambda^N \Lambda_\alpha
\;, 
\eea
with weight $\lambda$ under generalized diffeomorphisms, the particular combination
\bea
T_M &\equiv& (t^\alpha)_M{}^N\,\partial_N \Lambda_\alpha
\;,
\eea
of internal derivatives, does not transform tensorially under generalized diffeomorphisms, but rather as
\bea
\delta_\Lambda T_M &=& 
{\cal L}_\Lambda^{(\lambda-\lambda_d)}T_M
+ (t^\alpha)_M{}^N\,  (\lambda-1)\,\partial_N \partial_K \Lambda^K \Lambda_\alpha
+(t^\gamma)_L{}^K  \,\partial_M\partial_K\Lambda^L\,\Lambda_\gamma
\;,
\label{deltaT}
\eea
provided $Y^{MN}{}_{KL}\,(t_\gamma)_N{}^P\,\partial_P \partial_M  = 0$ which holds for all 
E$_{d(d)}$, $d<8$, as a consequence of the section constraint.
In particular, for $\lambda=1=(9-d)\,\lambda_d$, which is the correct ExFT weight for $(9-d)$-forms, the non-covariant
transformation of $T_M$ reduces to the last term of (\ref{deltaT}) and can be absorbed by
introducing a covariantly constrained object $\Xi_M$ to which one assigns the
transformation law
\bea\label{deltaXIII}
\delta_\Lambda \Xi_M&=&{\cal L}_\Lambda^{(1-\lambda_d)}\,\Xi_M 
-(t^\gamma)_L{}^K  \,\partial_M\partial_K\Lambda^L\,\Lambda_\gamma
\nonumber\\
&=&
\Lambda^K \partial_K \Xi_M 
+ \partial_M\Lambda^K\,\Xi_K
+\partial_K\Lambda^K\,\Xi_M
-(t^\gamma)_L{}^K  \,\partial_M\partial_K\Lambda^L\,\Lambda_\gamma
\;,
\eea
under generalized diffeomorphisms. `Covariantly constrained' indicates that this object satisfies
the same section constraint (\ref{section}) as the internal derivative operators, i.e.\
\bea
Y^{MK}{}_{NL}\,\Xi_M\, \partial_K &=& 0  \; = \;  Y^{MK}{}_{NL}\,\Xi_M\, \Xi_K
\;.
\label{sectionXi}
\eea

The resulting complete gauge transformation for ${\cal B}_M$ then extends (\ref{deltaBexp})  to
\bea
\delta_{\Lambda,\Xi} {\cal B}_M &=& D\Lambda_M + \dots - (t^\alpha)_M{}^N\,\partial_N \Lambda_\alpha - \Xi_M
\;,
\label{fullDeltaB}
\eea
with an $(8-d)$-form covariantly constrained gauge parameter $\Xi_M$. Its associated
$(9-d)$-form gauge field ${\cal C}_M$ then appears in the full covariant field strength as
\bea
{\cal F}_M &=& D {\cal B}_M + \dots + 
(t^\alpha)_M{}^N\,\partial_N {\cal C}_\alpha + {\cal C}_M
\;.
\label{Ffull}
\eea
We will give the explicit expressions for E$_{7(7)}$ and E$_{8(8)}$ ExFT in the next subsections.
W.r.t.\ the higher-dimensional origin of the ExFT fields, the presence of ${\cal C}_M$ in the field
strength ${\cal F}_M$ is precisely the remnant of the required presence of the compensating field $Y$
in the full non-linear field strength of the dual graviton, c.f.\ (\ref{DBF}).

Let us briefly describe the generic structure of the dynamics of the field ${\cal C}_M$.
The Bianchi identity for the field strength (\ref{Ffull}) takes the form
\bea
D {\cal F}_M &=& 
\star(J_{\rm CS})_M  +(t^\alpha)_M{}^N \partial_N {\cal F}_\alpha  +{\cal G}_M 
\;,
\label{Bianchi}
\eea
with the Chern-Simons contributions $\star(J_{\rm CS})_M$ resulting from the ellipsis in (\ref{Ffull})
and ${\cal F}_\alpha$ and ${\cal G}_M$ denoting the non-abelian field strengths of 
${\cal C}_\alpha$ and ${\cal C}_M$, respectively.
The dynamics of the $(8-d)$-forms is encoded in a first-order duality equation
\bea
{\cal F}_M &=& {\cal M}_{MN}\,\star{\cal F}^N
\;,
\label{dualMM}
\eea
where ${\cal M}_{MN}$ denotes the scalar dependent E$_{d(d)}$ matrix parametrizing the scalar target space,
and ${\cal F}^N$ represents the non-abelian field strength for the ExFT vector fields in the fundamental representation of E$_{d(d)}$. The second order field equations for the latter are derived from the ExFT Lagrangian and take the generic form
\bea
D\left( {\cal M}_{MN}\,\star{\cal F}^M \right)
&=& (J_{\rm CS})_M 
+(t^\alpha)_M{}^N\,\partial_N\,{\cal J}_\alpha
+ I_{M}
\;.
\label{YM}
\eea
Here, $(J_{\rm CS})_M$ comes from variation of the topological terms of the Lagrangian,
while ${\cal J}_\alpha$ and $I_{M}$ derive from variation of the vector fields within connections (\ref{covder}), 
with ${\cal J}_\alpha$ carrying the contributions from the rotation term and ${I}_M$ carrying
the contributions from the transport and the weight term. In particular, $I_M$ is covariantly
constrained according to the notion of (\ref{sectionXi}).

Combining exterior derivative of the duality equation (\ref{dualMM}) with the Bianchi 
identity (\ref{Bianchi}) and the Yang-Mills equations (\ref{YM}), shows that the CS terms cancel 
as they do in the dimensionally reduced theory \cite{Cremmer:1998px} (when $\partial_M=0$), 
such that we are left with the duality equations
\bea
{\cal F}_\alpha &=& \star {\cal J}_\alpha
\;,
\label{HH1}\\
{\cal G}_M &=& \star I_{M}
\;,\label{HH2}
\eea
describing the dynamics of the fields ${\cal C}_\alpha$ and ${\cal C}_M$ descending from the
higher-dimensional dual graviton and the compensating field $Y$, respectively. They encode
some of the components of the higher-dimensional duality equations 
(\ref{dualitygrav}) and (\ref{2ndduality}), respectively.

Finally, we would like to point out that ${\cal F}_\alpha$ and ${\cal G}_M$ 
(and likewise $\Lambda_{\alpha}$ and $\Xi_M$) naturally form an irreducible object 
w.r.t.~an underlying Lie algebra from which the algebra of 
generalized diffeomorphisms can be derived by means of an 
`embedding tensor'. Specifically, 
starting from the Lie algebra $\frak{g}$ of (conventional) diffeomorphisms and local U-duality transformations, 
spanned by parameters $(\lambda^M, \sigma^{\alpha})$, the generalized diffeomorphisms of ExFT can be defined 
in terms of $\frak{g}$-representations and the embedding tensor $\vartheta: R\rightarrow \frak{g}$, 
where $R$ denotes the representation labelled by indices $M, N$, as \cite{Hohm:2018ybo}
 \be
  \vartheta(\Lambda) \ = \ (\Lambda^M\,,\;-\kappa(t^{\alpha})_M{}^{N}\partial_N\Lambda^M) \;. 
 \ee
The objects ${\cal F}_{\alpha}$ and ${\cal G}_M$ form now the irreducible \textit{coadjoint} representation of $\frak{g}$ 
in that the transformations (\ref{deltaLambda}) and (\ref{deltaXIII}) can be rewritten 
in terms of the coadjoint action as 
 \be\label{coadjointREP}
  \delta_{\Lambda}\begin{pmatrix}   {\cal F}_{\alpha}  \\
  {\cal G}_M   \end{pmatrix} \ \equiv \ {\rm ad}_{\vartheta(\Lambda)}^* 
  \begin{pmatrix}   {\cal F}_{\alpha}  \\
  {\cal G}_M   \end{pmatrix}\;, 
 \ee
as one may quickly verify with eq.~(A.9) in \cite{Hohm:2018ybo}.

\subsection{E$_{7(7)}$ general structure and solving the section constraint}

In E$_{7(7)}$ ExFT, the fields with origin in the $11$-dimensional
dual graviton show up among the vector and the two-form fields. More
precisely, the fields (\ref{dualGRAVcomppp}) give rise to $7$ vectors $B_{\mu\,m}$
and $49$ two-forms $E_{\mu\nu,m,n}$, together with the $D=4$ dual graviton $C_{\mu\nu\rho,\sigma}$\,.
In this case, the field strength (\ref{Ffull}) in which the constrained compensator field first
appears is the field strength ${\cal F}^M= \Omega^{MN}{\cal F}_N$ of the vector fields itself
which is explicitly given by
\bea
{\cal F}_{\mu\nu}{}^M &= & 
2 \,\partial_{[\mu} {\cal A}_{\nu]}{}^M 
-2\,{\cal A}_{[\mu}{}^K \partial_K {\cal A}_{\nu]}{}^M 
-\frac1{2}\,
\Omega^{MP}
\left(24\, (t_\alpha)_P{}^{K} (t^\alpha)_{L}{}^Q
-
\delta_P^K\delta_L^Q
\right)
\Omega_{QN}
{\cal A}_{[\mu}{}^N\,\partial_K {\cal A}_{\nu]}{}^L
\nonumber\\
&&{}
- 12 \,  \Omega^{MK}\,(t^\alpha)_K{}^{N} \,\partial_N {\cal B}_{\mu\nu\,\alpha}
-\frac12\,\Omega^{MK}\,{\cal B}_{\mu\nu\,K}
\;,
\label{YM4}
\eea
in terms of the E$_{7(7)}$ generators $(t_\alpha)_M{}^{N}$ and the symplectic matrix $\Omega_{MN}$\,.
The last two terms in (\ref{YM4}) correspond to the couplings introduced in (\ref{Ffull}) with two-form gauge potentials ${\cal B}_{\mu\nu\,\alpha}$ and ${\cal B}_{\mu\nu\,M}$ in the adjoint and the fundamental representation,
respectively.

In $D=4$ ExFT (and supergravity), the generic duality equation (\ref{dualMM}) is replaced by the
twisted self-duality equations~\cite{Cremmer:1998px}
\bea
{\cal F}_{\mu\nu}{}^M &=& -\frac12\,\sqrt{|g|}\,\varepsilon_{\mu\nu\rho\sigma}\,\Omega^{MN}{\cal M}_{NK}{\cal F}^{\rho\sigma\,K}
\;.
\label{duality56}
\eea
It is accompanied by the duality equations between two-forms and scalar fields (\ref{HH1}), (\ref{HH2}),
which here take the explicit form
\bea
 {\cal H}_{\mu\nu\rho\,\alpha} &=&
-\sqrt{|g|}\, \varepsilon_{\mu\nu\rho\sigma}\,
(t_\alpha)_K{}^L
\left({D}^\sigma {\cal M}^{KP} {\cal M}_{LP} \right)
\;,\nonumber\\
{\cal H}_{\mu\nu\rho\,M}&=&
\sqrt{|g|}\, \varepsilon_{\mu\nu\rho\sigma}\left(
\widehat{J}^\sigma{}_{M} -\frac{1}{12}\,{ D}^\sigma {\cal M}^{KL} \partial_M {\cal M}_{KL} 
\right)
 \;,
 \label{dualH}
\eea
see~\cite{Hohm:2013uia} for details.

The ExFT section constraint is solved by decomposing the adjoint representation of E$_{7(7)}$ 
under its maximal GL$(7)$ subgroup and restricting internal derivatives according to 
 \be
  \partial_M \ = \ (\partial_m, 0,\ldots, 0)\;.
 \ee
The same decomposition applies to vector and two-forms and implies
\bea
{\cal A}_\mu{}^M &\longrightarrow& (
{\cal A}_\mu{}^m, {\cal A}_{\mu\,mn}, {\cal A}_\mu{}^{mn}, {\cal B}_{\mu\,m})\;,
\nonumber\\
{\cal B}_{\mu\nu\,\alpha} &\longrightarrow& 
({\cal B}_{\mu\nu\,m}, {\cal B}_{\mu\nu}{}^{kmn}, {\cal E}_{\mu\nu\,m}{}^n, \dots )
\;,
\nonumber\\
{\cal B}_{\mu\nu\,M} &\longrightarrow& (
{\cal C}_{\mu\nu\,m},0,0,0)
\;,
\label{vec4}
\eea 
where we have restricted to those fields actually appearing in (\ref{duality56})
and made explicit the fields ${\cal B}_{\mu\,m}$, ${\cal E}_{\mu\nu\,m}{}^n$, ${\cal C}_{\mu\nu\,m}$,
descending from the $D=11$ dual graviton (${\cal B}_{\mu\,m}$, ${\cal E}_{\mu\nu\,m}{}^n$) 
and the compensator field (${\cal C}_{\mu\nu\,m}$), respectively.
Zooming in on their field equations, the relevant component of the field strength (\ref{YM4})
gives the full completion of (\ref{Fm4}) as
\bea
{\cal F}_{\mu\nu\,m} &=&
2\,\partial_{[\mu} {\cal B}_{\nu]\,m}
-2\,{\cal A}_{[\mu|}{}^k \partial_k {\cal B}_{|\nu] m}
-2\,\partial_m {\cal A}_{[\mu}{}^k  {\cal B}_{\nu] k}
-2\,\partial_k {\cal A}_{[\mu}{}^k  {\cal B}_{\nu] m}
\nonumber\\
&&{}
+2\,\partial_{k} {\cal A}_{[\mu}{}^{kl} {\cal A}_{\nu] lm}
-3\,{\cal A}_{\mu}{}^{kl} \partial_{[m} {\cal A}_{|\nu| kl]}
+\partial_n {\cal E}_{\mu\nu,m}{}^n-\partial_m {\cal E}_{\mu\nu,n}{}^n +{\cal C}_{\mu\nu\,m}
\;.
\label{Fm7}
\eea
By virtue of (\ref{duality56}), this field strength is expressed as the dual of the remaining
field strengths, reproducing the relevant components of the duality equations (\ref{dualitygrav}).
In turn, the second equation of (\ref{dualH}) reproduces the corresponding components of the
eleven-dimensional equation (\ref{2ndduality}) for the compensator field.
The combination of (\ref{duality56}) and (\ref{dualH}) thus reproduces the correct eleven-dimensional dynamics.
Comparing the explicit form of the field strength (\ref{Fm7}) to its higher-dimensional ancestor (\ref{shiftfield}),
the first line of (\ref{Fm7}) descends from the first term of  (\ref{shiftfield}) upon expanding the objects 
in the standard Kaluza-Klein basis (see e.g.\ \cite{Hohm:2013vpa} for a detailed discussion).
The ${\cal A}^2$ terms in the second line of (\ref{Fm7}) indicate the translation of the relevant
component ${\tilde Y}_{\mu\nu\,m}$ of the higher-dimensional compensator field into the
component ${\cal C}_{\mu\nu\,m}$ of the duality covariant ExFT object ${\cal B}_{\mu\nu\,M}$ (\ref{vec4}).

This illustrates  how the E$_{7(7)}$ covariant ExFT field equations (\ref{duality56}), (\ref{dualH}) require the inclusion of
certain components of the eleven-dimensional equations (\ref{dualitygrav}), (\ref{2ndduality}) 
for the dual graviton and the compensator field, respectively,
which in turn become part of the duality covariant ExFT objects as in (\ref{vec4}).

\subsection{E$_{8(8)}$ general structure and solving the section constraint}

We now spell out the details of the general structures introduced above for the E$_{8(8)}$ case. 
The adjoint representation of E$_{8(8)}$ is 248-dimensional, and we denote its indices by $M,N=1\,\ldots, 248$
 and the  
structure constants 
by $f^{MN}{}_{K}$. The tensor product ${\bf 248}\otimes {\bf 248}$ 
decomposes as 
 \be\label{248squared}
  {\bf 248}\otimes {\bf 248} \ \rightarrow \ {\bf 1}\oplus  {\bf 248}\oplus {\bf 3875}\oplus {\bf 27000}
  \oplus {\bf 30380}\;, 
 \ee 
and the section constraints project out the  sub-representation ${\bf 1}\oplus  {\bf 248}\oplus {\bf 3875}$, 
i.e., 
   \be\label{secconstr}
  \eta^{MN}\partial_M\otimes  \partial_N \ = \ 0 \;, \quad 
  f^{MNK}\partial_N\otimes \partial_K \ = \ 0\;, \quad 
  (\mathbb{P}_{3875})_{MN}{}^{KL} \partial_K\otimes \partial_L \ = \ 0
  \;. 
 \ee
The explicit form of the projectors can be found in \cite{Hohm:2014fxa}. 

In E$_{8(8)}$ ExFT, the fields on which additional gauge transformations have to be introduced as in 
(\ref{fullDeltaB}) are the scalar fields.
As a result, these transformations modify the generalized diffeomorphisms (\ref{gen_diff}) themselves
which become
 \be\label{usualgenLie}
  {\cal L}^{[\lambda]}_{(\Lambda,\Sigma)}V^M \ = \ \Lambda^N\partial_N V^M 
  +f^{M}{}_{NK} R^N V^K +\lambda\, \partial_N\Lambda^N V^M\;, 
 \ee
where 
 \be\label{RDEF}
  R^M \ \equiv \ f^{MN}{}_{K}\,\partial_N\Lambda^K +\Sigma^M\;.   
 \ee
This expression is the E$_{8(8)}$ implementation of the general structure (\ref{fullDeltaB}) discussed above, combining 
the gauge parameter of $(8-d)$-forms (i.e., scalars for the E$_{8(8)}$ case) with a 
`covariantly constrained' parameter in the adjoint representation (i.e., here in the fundamental representation of E$_{8(8)}$). 
As shown recently in \cite{Hohm:2017wtr,Hohm:2018ybo}, the space of these extended gauge parameters, 
which we group as $\Upsilon = \big(\Lambda^M, \Sigma_M\big)$, carries the structure of a Leibniz-Loday algebra 
with product 
  \be\label{LeibnizE8Prod}
  \Upsilon_1 \circ  \Upsilon_2 \ \equiv \ \Big(\,{\cal L}_{\Upsilon_1}^{[1]}\Lambda_2{}^M \;,\; \,
  {\cal L}_{\Upsilon_1}^{[0]}\Sigma_{2M} \ + \ \Lambda_2{}^N\partial_M R_{N}(\Upsilon_1)\,\Big)\;, 
 \ee 
which satisfies the Leibniz rule 
 \be\label{Loday}
  \Upsilon_1\circ (\Upsilon_2\circ \Upsilon_3) \ = \ (\Upsilon_1\circ \Upsilon_2) \circ \Upsilon_3 
  + \Upsilon_2\circ (\Upsilon_1\circ \Upsilon_3)\;. 
 \ee
Thanks to this algebraic structure, the E$_{8(8)}$ ExFT can be formulated efficiently in terms 
of `doubled' parameters and fields,\footnote{This doubled structure can also be understood as a special case of the general realization in terms of the coadjoint action of a Lie algebra in (\ref{coadjointREP}), using 
that here the adjoint and $R$ representations coincide.} 
which makes manifest that the consistency of the theory hinges on 
the precise interplay of the naive (generalized diffeomorphism) parameter $\Lambda^M$ and the 
new companion parameter $\Sigma_M$ required for consistency of the dual graviton couplings, 
as we will explain in more detail in the following.

In order to explain the resolution of the dual graviton problem for the E$_{8(8)}$ case more explicitly, 
we first  review the bosonic field content. It is given by the external dreibein $e_{\mu}{}^{a}$, 
carrying curved and flat 3D indices, the internal 248-bein ${\cal V}_{A}{}^{M}$, being a matrix in 
the adjoint of E$_{8(8)}$, and gauge vectors ${\frak A}_{\mu}\equiv ({\cal A}_{\mu}{}^{M}, {\cal B}_{\mu M})$ 
taking values in the Leibniz-Loday algebra. All fields depend on $3+248$ coordinates $(x^{\mu}, Y^M)$,
modulo the section constraints. 
The action takes the schematic form 
  \be
  S \ = \ \int d^3x \, d^{248}Y \,\sqrt{|g|}\left( \widehat{R}+\sqrt{|g|}{}^{-1}{\cal L}_{\rm CS}
  +\frac{1}{240}\,g^{\mu\nu}
  { D}_{\mu}{\cal M}^{MN}{ D}_{\nu}{\cal M}_{MN} -\,V({\cal M},g) \right)\;. 
  \label{fullaction0}
 \ee 
The first term is the 3D Einstein-Hilbert term for  $e_{\mu}{}^{a}$, suitably covariantized w.r.t.~the 
gauge connection ${\frak A}_{\mu}$. The second term is a Chern-Simons term for ${\frak A}_{\mu}$ 
based on the Leibniz-Loday algebra structure (\ref{Loday}) \cite{Hohm:2017wtr,Hohm:2018ybo}. 
The third term is the kinetic term for the E$_{8(8)}/{\rm SO}(16)$ coset 
scalar matrix ${\cal M}\equiv {\cal V}{\cal V}^{\rm T}$. 
The final term is the `potential' (in the sense that it involves only external derivatives and hence 
reduces upon Kaluza-Klein reduction to a genuine scalar potential),  
 \bea\label{PotIntro}
  V({\cal M},g) & = &
  -\frac{1}{240}{\cal M}^{MN}\partial_M{\cal M}^{KL}\,\partial_N{\cal M}_{KL}+
  \frac{1}{2}{\cal M}^{MN}\partial_M{\cal M}^{KL}\partial_L{\cal M}_{NK} \\
  &&{}
  +\frac1{7200}\,f^{NQ}{}_P f^{MS}{}_R\,
  {\cal M}^{PK} \partial_M {\cal M}_{QK} {\cal M}^{RL} \partial_N {\cal M}_{SL} 
   \ + \ \text{$(\partial g)$ terms}\;, 
    \nonumber
\eea
where we suppressed terms carrying derivatives of the external metric $\partial_M g_{\mu\nu}$,
see \cite{Hohm:2014fxa} for the full expression. 
Upon solving the section constraint by decomposing the adjoint representation of E$_{8(8)}$ 
under its maximal GL$(8)$ subgroup and 
writing for the internal derivatives 
 \be
  \partial_M \ = \ (\partial_m, 0,\ldots, 0)\;, 
 \ee
where $m=1,\ldots, 8$ is the fundamental GL$(8)$ index, 
the above action is fully equivalent to $D=11$ supergravity in a $3+8$ split.

Let us now discuss this match with $D=11$ supergravity schematically, focusing on the role of 
dual fields such as the dual graviton. Decomposing the ExFT fields under the maximal GL$(8)$ subgroup
that survives as a manifest symmetry, one obtains the following fields: 
The dreibein $e_{\mu}{}^{a}$, being an E$_{8(8)}$ singlet, does not decompose and originates 
directly from the corresponding 3D component of the vielbein in $D=11$ under the $3+8$ split. 
In particular, there is no `external dual graviton'. The E$_{8(8)}/{\rm SO}(16)$ coset degrees of freedom 
can be parameterized in terms of the following GL$(8)$ covariant scalar fields: 
 \be\label{scalardecomp}
  {\cal V}_A{}^M\;:\qquad  \phi_{mn} \ \in \ {\rm GL}(8)/{\rm SO}(8)\;, \quad 
  c_{mnk}\;, \quad c_{m_1\ldots m_6}\;, 
  \quad  c_{n_1\ldots n_8,m} \ \equiv \ \varepsilon_{n_1\dots n_8}\,\varphi_m\;. 
 \ee
Here $\phi_{mn}$ is symmetric and originates directly as `Kaluza-Klein scalars' 
from the $D=11$ frame field, as in (\ref{KKsplit}).  
The fields denoted by $c$ are completely antisymmetric and thus naturally originate as internal components 
from a 3-form and 6-form  in $D=11$. Finally, the field $\varphi_m$ is equivalent, 
as indicated here, to an $[8,1]$ field, thus suggesting an interpretation as a dual graviton component.
Next, the vector fields decompose into GL$(8)$ covariant fields as follows: 
 \be\label{vectordecomp}
  \begin{split}
   &{\cal A}_{\mu}{}^{M}\,:\qquad {\cal A}_\mu{}^m\,, \quad {\cal A}_\mu{\,}_{mn}\,, \quad {\cal A}_\mu{\,}_{mnklp}\,, \quad 
   {\cal A}_\mu{\,}_m{}^n\,, \quad \cdots \\
   & {\cal B}_{\mu M}\,:  \qquad {\cal B}_{\mu m}\,, \quad \cdots \;. 
  \end{split}
 \ee  
Here we have left out a number of vector fields, indicated by the ellipsis, that drop out in the final action upon 
solving the section constraint. In particular, among the covariantly constrained vector fields ${\cal B}_{\mu M}$ 
only eight components survive, c.f.~(4.3), (4.4) in \cite{Hohm:2014fxa}. 

We are now ready to account for the $D=11$ fields under the $3+8$ split. The $D=11$ frame field 
yields a 3D frame field and an internal ${\rm GL}(8)/{\rm SO}(8)$ coset matrix, both of which 
were already identified among the above fields. It also gives rise to eight Kaluza-Klein vectors ${\cal A}_{\mu}{}^{m}$
that are among the components of ${\cal A}_{\mu}{}^{M}$ in (\ref{vectordecomp}). 
However, here we seem to encounter the dual graviton problem, because naively the scalars $\varphi_m$
in (\ref{scalardecomp}) also encode the degrees of freedom of the Kaluza-Klein vectors. 
Indeed, in dimensional reduction it is necessary to dualize the Kaluza-Klein vectors ${\cal A}_{\mu}{}^{m}$
into 3D scalars $\varphi_m$ in order to complete the E$_{8(8)}/{\rm SO}(16)$ coset matrix. 
Since in ExFT we have both ${\cal A}_{\mu}{}^{m}$ and $\varphi_m$ there seems to be an over-counting 
of fields. In order to see that, on the contrary, the degrees of freedom properly match 
we have to take into account the compensating gauge vectors ${\cal B}_{\mu m}$. 
The Chern-Simons and scalar kinetic terms then take the schematic form 
 \be\label{actionscalarr}
  {\cal L}_{\rm scalar-vector} \ = \ \tfrac{1}{2}\, \varepsilon^{\mu\nu\rho} {\cal B}_{\mu m} {\cal F}_{\nu\rho}{}^{m} - 
  \tfrac{1}{2}\,\phi^{mn} D^{\mu}\varphi_m D_{\mu}\varphi_n\;, 
 \ee
where ${\cal F}_{\mu\nu}{}^{m}$ is the field strength for the Kaluza-Klein vectors and 
 \be
  D_{\mu}\varphi_m \ = \ \partial_{\mu}\varphi_m \ - \ {\cal L}_{{\cal A}_{\mu}}\varphi_m  \ + \  {\cal B}_{\mu m}\;. 
 \ee  
We observe here St\"uckelberg couplings, in agreement with the residual gauge invariance 
originating from the $\Sigma_M$ transformations, $\delta\varphi_m=-\Sigma_m$, 
$\delta {\cal B}_{\mu m}=D_{\mu}\Sigma_m$. Consequently, upon integrating out ${\cal B}_{\mu m}$ from (\ref{actionscalarr}) 
the scalar components $\varphi_m$ drop out, and the action reduces to the Yang-Mills term for the 
vectors ${\cal A}_{\mu}{}^{m}$, in precise agreement with the 3+8 split of $D=11$ supergravity. 
Moreover, working out the explicit form of the covariant derivatives (\ref{covder}) with (\ref{usualgenLie})
shows that the vector components ${\cal A}_\mu{\,}_m{}^n$ (which are the 
three-dimensional version of the $(n-2)$-forms ${E}_{m,n}$ discussed in sec.~2.2)  
only enter in the combination $\partial_n {\cal A}_{\mu m}{}^{n}+{\cal B}_{\mu m}$, so that integrating out 
${\cal B}_{\mu m}$ also eliminates ${\cal A}_\mu{\,}_m{}^n$ from the Lagrangian.
In addition, in the scalar potential (\ref{PotIntro}) the components $\varphi_m$ drop out after solving 
the section constraint, as it should be in view of the St\"uckelberg invariance. 
Thus, there is no conflict with the presence of the `dual graviton' field $\varphi_m$ in ExFT and 
the fact that it can be matched with $D=11$ supergravity without a dual graviton field. 

Finally, let us identify the degrees of freedom corresponding to the 3-form in $D=11$, 
which gives rise to scalars $c_{mnk}$, vectors ${\cal A}_{\mu mn}$, 2-forms ${\cal B}_{\mu\nu m}$ and a 
3-form ${\cal C}_{\mu\nu\rho}$. The scalars and vectors are already contained in (\ref{scalardecomp}) 
and (\ref{vectordecomp}), respectively. The 2-forms are not present explicitly in ExFT but rather 
encoded in the scalars $c_{m_1\ldots m_6}$ in (\ref{scalardecomp}) via the duality relation 
 \bea
H_{\mu\nu\rho\,m} \ = \ e\,\varepsilon_{\mu\nu\rho}\,\epsilon_{m n_1 \dots n_7} \,F^{n_1 \dots n_7} 
\;,
\label{Fj3}
\eea
where $H_m$ is the field strength of the 2-form, and 
$F_{n_1 \dots n_7} =7\,\partial_{[n_1} c_{n_2 \dots n_7]}+140\,c_{[n_1n_2n_3} \partial_{n_4} c_{n_5n_6n_7]}$. 
Next, the 4-form curvature of the 3-form vanishes identically in 3D and can be eliminated from the action 
(using techniques similar to those in \cite{Hohm:2013vpa}). 
Thus, all fields originating from the  3-form in $D=11$ are accounted for. 
Finally, the vectors ${\cal A}_\mu{\,}_{mnklp}$ in (\ref{vectordecomp}) are defined in terms of 
the scalars $c_{m_1\ldots m_6}$, which were introduced in (\ref{Fj3}), via 
 \be
  F_{\mu\nu\,m_1\ldots m_5} \ = \ e\,\varepsilon_{\mu\nu\rho}\,\epsilon_{m_1 \dots m_5 nkl}\,{\cal J}^{\rho\, nkl}\;, 
 \ee
with the current ${\cal J}_{\mu}{}^{mnk}=\epsilon^{mnk p_1\ldots p_5}\,\partial_{\mu}c_{p_1\ldots p_5}+\cdots $,
with the ellipsis indicating the connection term.

\section{Application: Reductions and consistent truncation on AdS$_4\times\, ${\rm S}$^7$}

Exceptional field theory provides a powerful tool to study consistent truncations of maximal
supergravity by virtue of a generalized Scherk-Schwarz reduction. 
These are truncations to lower-dimensional theories, such that every solution to the lower-dimensional
field equations lifts to a solution of the higher-dimensional field equations. In particular, this requires that
all dependence on the internal coordinates factors out from the higher-dimensional field equations.
In this section, we work out the reduction formulas for the dual graviton and the
compensator field for the prominent example of the truncation of $D=11$ supergravity on 
AdS$_4\times {\rm S}^7$~\cite{deWit:1986iy}.
In turn, this illustrates the necessity of the ExFT compensator field in order to consistently reproduce 
part of the higher-dimensional equations for the dual graviton around the seven-sphere.

The reduction formulas for the $p$-form fields
in a consistent Scherk-Schwarz type reduction of E$_{7(7)}$ ExFT 
read \cite{Hohm:2014qga}
  \bea\label{SchSch}
  {\cal A}_{\mu}{}^{M}(x,Y) &=& \rho^{-1}(Y) \,(U^{-1})_{N}{}^{M}(Y)\,A_{\mu}{}^{N}(x) \;, 
  \nonumber\\
  {\cal B}_{\mu\nu\,\alpha}(x,Y) &=& \,\rho^{-2}(Y)\, U_\alpha{}^\beta(Y)\,B_{\mu\nu\,\beta}(x)
  \nonumber\\
{\cal B}_{\mu\nu\,M}(x,Y) &=&
-2\, \rho^{-2}(Y)\,(U^{-1})_S{}^P(Y) \,\partial_M U_P{}^R(Y) (t^{\alpha}){}_R{}^S\, B_{\mu\nu\,\alpha}(x) 
  \;,
 \eea
in terms of an E$_{7(7)}$ valued twist matrix $U$ in the fundamental representation,
together with a scaling factor $\rho$. For the $S^7$ reduction, the twist matrix $U$ lives 
in the subgroup ${\rm SL}(8)\subset {\rm E}_{7(7)}$ and is of the explicit form
$U_{\underline{m}}{}^a = \{U_0{}^a, U_m{}^a\}$ with
\bea
U_0{}^a &=& \omega^{3/4}\,{\cal Y}^a 
-6\,\omega^{-1/4}\,\zeta^n \partial_n {\cal Y}^a
\;,
\nonumber\\
U_m{}^a &=& \omega^{-1/4}\,\partial_m {\cal Y}^a
\;,
\eea
in terms of sphere harmonics ${\cal Y}^a{\cal Y}^a=1$ and $\partial_n\zeta^n = \omega \equiv \sqrt{|g_{S^7}|}$,
\cite{Lee:2014mla,Hohm:2014qga}.
The scale factor $\rho$ is given by $\rho=\omega^{-1/2}$\,.

For the different components (\ref{vec4}) of the ExFT vector fields, 
the reduction formula from (\ref{SchSch}) then implies the explicit reduction
formulas
\bea
&& {\cal A}_\mu{}^m 
= 
{\cal K}^m{}_{ab}\,A_\mu{}^{ab}
 \;,\qquad
 {\cal A}_{\mu}{}^{mn} =\left(
\omega\, {\cal K}^{mn}{}_{ab}+
12\,\zeta^{[m} {\cal K}_{ab}{}^{n]} 
\right)
\,A_\mu{}^{ab}
 \;,\nonumber\\
&& {\cal A}_{\mu\,mn} =
{\cal K}_{mn}{}^{ab}\,A_{\mu\,ab}
 \;,\qquad
 {\cal B}_{\mu\,m} 
  =
 \left(
\omega\,{\cal K}_m{}^{ab} 
+6\,\zeta^n {\cal K}_{mn}{}^{ab}  \, 
\right)
  A_{\mu\,ab}
  \;,
  \label{red_vec}
\eea
with $\{ A^{ab}, A_{ab}\}=\{A^M\}$ denoting the 28 electric and 28 magnetic vector fields of $D=4$ supergravity, 
and the $S^7$ Killing vectors and tensors defined by
\bea
{\cal K}^{ab}{}_m \equiv  {\cal Y}^{[a} \partial_m {\cal Y}^{b]}\;,\qquad
K_{mn}{}^{ab} \equiv
\partial_{[m} {\cal K}^{ab}{}_{n]} 
\;.
\eea
This reproduces the formulas from \cite{deWit:1986iy,deWit:2013ija,Godazgar:2013pfa}.
In particular, the last formula of (\ref{red_vec}) provides the reduction formula for the 
components $B_{\mu\,m}$ from the 11-dimensional dual graviton (\ref{dualGRAVcomppp}).

The other components ${\cal E}_{\mu\nu,m}{}^n$ of the 11-dimensional dual graviton are 
identified within the ExFT two-forms according to (\ref{vec4}). The relevant reduction formula (\ref{SchSch})
then implies its reduction as
\bea
{\cal E}_{\mu\nu,m}{}^n &=&
\left(
6\,\partial_m {\cal Y}^a {\cal Y}_b\, \zeta^n  +  \sqrt{ |g_{S^7}|}\,\partial_m {\cal Y}^a \, \partial^n {\cal Y}_b\right)
\,B_{\mu\nu\,a}{}^b
\;,
\label{red_E}
\eea
in terms of 63 of the 4-dimensional two-forms $\{B_{\mu\nu\,a}{}^b\} \subset \{B_{\mu\nu\,\alpha}\}$\,.
Finally, the components ${\cal C}_{\mu\nu\,m}$ from the $D=11$ compensator field sit within 
the constrained ExFT field ${\cal B}_{\mu\nu\,M}$ according to (\ref{vec4}) such that the last equation of (\ref{SchSch})
implies the following reduction formula
\bea
{\cal C}_{\mu\nu\,m} &=&
 -\Big(
\omega\,\partial^n {\cal Y}_b  \, \partial_m \partial_n {\cal Y}^a
+ \omega\, {\cal Y}_b  \,\partial_m {\cal Y}^a 
+\partial_m\omega \,  {\cal Y}_b \,{\cal Y}^a 
-6\, \partial_m  \zeta^n \,{\cal Y}_b\,\partial_n {\cal Y}^a 
  \Big)
\,B_{\mu\nu,a}{}^b
\;.
\label{red_C}
\eea
As discussed above, the eleven-dimensional duality equation for the dual graviton features
the non-abelian field strength (\ref{Fm7}). Consistency thus crucially requires that the full
non-linear expression (\ref{Fm7}) is compatible with the reduction formulas (\ref{red_vec})--(\ref{red_C}),
such that all dependence on the sphere coordinates factors out precisely as in the first term
\bea
\partial_{[\mu} {\cal B}_{\nu]\,m} 
  &=&
 \left(
\omega\,{\cal K}_m{}^{ab} 
+6\,\zeta^n {\cal K}_{mn}{}^{ab}  \, 
\right)
\partial_{[\mu}   A_{\nu]\,ab}
\;.
\label{redB1}
\eea
Collecting all the ${\cal A}{\cal B}$ and ${\cal A}^2$ terms from (\ref{Fm7}), it is lengthy but straightforward to show that
the various reduction formulas from (\ref{vec4}) indeed combine into
\bea
{\cal A}{\cal B}+{\cal A}{\cal A} &\longrightarrow& 
\left(
\omega\,{\cal K}_m{}^{ab} 
+6\,\zeta^n {\cal K}_{mn}{}^{ab}  \, 
\right)
A_{[\mu}{}^{ac} A_{\nu]}{}^{bc} 
\;,
\label{redB2}
\eea
i.e.\ complete (\ref{redB1}) into the non-abelian Yang-Mills field strength $F_{\mu\nu\,ab}$
for the magnetic vector fields.

It remains to analyze the two-form contributions in (\ref{Fm7}). The dual graviton contributions from
$\partial_n {\cal E}_{\mu\nu,m}{}^n$ are computed from (\ref{red_E}) as
\bea
\partial_n {\cal E}_{\mu\nu,m}{}^n-\partial_m {\cal E}_{\mu\nu,n}{}^n
&=&
\Big(
12\,{\cal K}_{mn}{}^a{}_b\, \zeta^n
-6\,\partial_n {\cal Y}^a {\cal Y}_b\, \partial_m\zeta^n 
+  \omega\,\partial_n \partial_m {\cal Y}^a \, \partial^n {\cal Y}_b
\nonumber\\
&&{}
\qquad
+\partial_m\omega \, {\cal Y}^a \,  {\cal Y}_b
+\omega\, {\cal Y}^a \, \partial_m {\cal Y}_b
\Big)
\,B_{\mu\nu,a}{}^b
\;,
\eea
i.e.~it is not consistent with the factorized form of (\ref{redB1}), (\ref{redB2}), unless the contribution (\ref{red_C})
from the compensator field is taken into account.
Upon combining all contributions from (\ref{red_vec})--(\ref{red_C}), the final result is
\bea
{\cal F}_{\mu\nu\,m} &=& 
\left(
\omega\,{\cal K}_m{}^{ab} 
+6\,\zeta^n {\cal K}_{mn}{}^{ab}  \, 
\right)
\left( 
F_{\mu\nu\,ab} + 2\,B_{\mu\nu,a}{}^{b}
\right)
\;.
\eea
The field strength (\ref{Fm7}) thus reduces in factorized form with the combination $F_{\mu\nu\,ab} + 2\,B_{\mu\nu,a}{}^{b}$
precisely capturing the St\"uckelberg corrected magnetic field strengths of maximal $D=4$ gauged supergravity~\cite{deWit:2007mt}.
As a result, the eleven-dimensional duality equation (\ref{dualitygrav}) for the dual graviton 
also allows for a consistent truncation on the seven-sphere.

\section{Summary and Outlook}

We reviewed the status of the so-called `dual graviton', with a particular emphasis on its role within 
duality covariant formulations 
of (the low-energy actions of) string/M-theory. There is no sharp physical problem 
that would require a field theory of a dual graviton, since one may always choose physical or light-cone 
gauge for which there is no difference between `graviton' and `dual graviton'. Nevertheless, 
in order to make certain features (such as duality symmetries) manifest, it is necessary to 
work with field variables that include dual graviton components. The main take-home message 
of this review should then be this: to the extent that duality-covariant formulations are the 
dual graviton's purpose in life there is a fully satisfactory 
formulation.

The formulation of `covariantly constrained' compensator gauge fields is the key ingredient 
that allowed us to circumvent the no-go theorems for the dual graviton. Here we summarize several results and 
observations that independently confirm that the seemingly bizarre notion of covariantly constrained fields 
and symmetries is self-consistent and necessary: 
 \begin{itemize}
 
  \item[(1)] The covariantly constrained fields and symmetries arise for any ExFT with duality group 
  E$_{d(d)}$, at the appropriate level of the tensor hierarchy of $p$-form potentials, to render the 
  dual graviton pure gauge. In particular, for the recently discussed generalized diffeomorphisms for the 
  affine E$_{9(9)}$ such fields are present already among the scalar fields \cite{Bossard:2017aae}.
 
  \item[(2)] The dynamical equations for the covariantly constrained fields are necessary in
  order to recover the correct eleven-dimensional dynamics from the duality covariant ExFT field equations.
 
  \item[(3)] In supersymmetric versions of ExFT, the covariantly constrained fields receive their 
  own independent supersymmetry variations \cite{Godazgar:2014nqa,Baguet:2016jph}, which are indispensable
  for closure of the supersymmetry algebra.

  \item[(4)] The consistency of generalized Scherk-Schwarz compactifications  requires the inclusion of 
  the compensator fields with an appropriate Scherk-Schwarz ansatz (that is manifestly compatible 
  with the constraints on the new fields). 
  These fields emerge already in consistency proofs of fully geometric compactifications such as 
  $D=11$ supergravity on AdS$_4\times {\rm S}^7$, as reviewed here. 
 
  \item[(5)] The doubled structure of gauge symmetries and vectors for the E$_{8(8)}$ theory
   has recently been shown to have a deeper mathematical significance in that, say, 
   the doubled vectors $\frak{A}_{\mu}\equiv (A_{\mu}{}^{M}, B_{\mu M})$ can be seen as 
   gauge vectors for so-called Leibniz-Loday algebras with a corresponding Chern-Simons formulation 
   of the action \cite{Hohm:2017wtr,Hohm:2018ybo}. 
 
 \end{itemize}

Finally, we point out that the above of course  does not exclude the possibility 
that some future applications may require a formulation containing dual graviton-type fields  
going beyond those discussed here. For instance, one may imagine that eventually a formulation 
is called for in which all mixed-Young tableaux fields are encoded in representations of the 
full duality group, most likely along the lines already investigated 
for double field theory at the linearized level \cite{Bergshoeff:2016ncb,Bergshoeff:2016gub}.

\section*{Acknowledgments} For useful comments and discussions and collaborations on related topics 
we would like to thank Eric Bergshoeff, Nicolas Boulanger, Chris Hull, and Hermann Nicolai.


\providecommand{\href}[2]{#2}\begingroup\raggedright\endgroup

\end{document}